# Highly Sensitive and Label-free Digital Detection of Whole Cell E. coli with Interferometric Reflectance Imaging

**Negin Zaraee [a], Fulya Ekiz kanik [a], Abdul Muyeed Bhuiya [b,c], Emily S. Gong [d], Matthew T. Geib [a], Nese Lortlar Ünlü [e], Ayca Yalcin Ozkumur [a,f], Julia R. Dupuis [d], M. Selim Ünlü [a,e,*]**

[a] Department of Electrical and Computer Engineering, Boston University, 8 St. Mary's Street, Boston, MA 02215, USA
[b] Department of Bioengineering, University of California, Berkeley, CA 94720, USA
[c] Department of Bioengineering and Therapeutic Sciences, University of California, San Francisco, CA 94158, USA
[d] Physical Science Inc., 20 New England Business Center, Andover, MA 01810, USA
[e] Department of Biomedical Engineering, Boston University, 44 Cummington Mall, Boston, MA 02215, USA
[f] Department of Electrical and Electronics Engineering, Bahçeşehir University, Istanbul, Turkey

## Abstract

Bacterial infectious diseases are a major threat to human health. Timely and sensitive pathogenic bacteria detection is crucial in identifying the bacterial contaminations and preventing the spread of infectious diseases. Due to limitations of conventional bacteria detection techniques there have been concerted research efforts towards development of new biosensors. Biosensors offering label-free, whole bacteria detection are highly desirable over those relying on label-based or pathogenic molecular components detection. The major advantage is eliminating the additional time and cost required for labeling or extracting the desired bacterial components. Here, we demonstrate rapid, sensitive and label-free *E. coli* detection utilizing interferometric reflectance imaging enhancement allowing for visualizing individual pathogens captured on the surface. Enabled by our ability to count individual bacteria on a large sensor surface, we demonstrate a limit of detection of 2.2 CFU/ml from a buffer solution with no sample preparation. To the best of our knowledge, this high level of sensitivity for whole *E. coli* detection is unprecedented in label-free biosensing. The specificity of our biosensor is validated by comparing the response to target bacteria *E. coli* and non-target bacteria *S. aureus*, *K. pneumonia* and *P. aeruginosa*. The biosensor's performance in tap water also proves that its detection capability is unaffected by the sample complexity. Furthermore, our sensor platform provides high optical magnification imaging and thus validation of recorded detection events as the target bacteria based on morphological characterization. Therefore, our sensitive and label-free detection method offers new perspectives for direct bacterial detection in real matrices and clinical samples.

## 1. Introduction

Bacterial infections and contaminations are a serious threat to public health, claiming millions of lives every year. The mortalities from bacterial infections are particularly high in developing countries among more vulnerable populations including children, elderly and patients with poor medical conditions. Contaminated food and water are one of the most frequent causes of bacterial infections and their rapid spread among consumers. As reported by World Health Organization (WHO) in a 2015 study, consuming contaminated food causes around 550 million people– almost 1 in 10 people in the world – to fall ill and an estimated 420,000 fataozlities annually (WHO 2015). In addition, waterborne diarrheal diseases are estimated to be responsible for 2 million deaths each year, with the majority occurring in children under 5 years old (WHO). *Escherichia coli* is one of the five most common bacterial contaminants, not only in developing countries, but also in developed countries such as US and Canada. According to a 2018 report published by Centers for Disease Control and Prevention (CDC) there have been more than 10 multistate *E. coli* outbreaks in US only from 2014 to 2018 (CDC). *E. coli* infection can lead to bloody diarrhea and if not treated immediately, to kidney failure and death (Nordqvist 2017). Therefore, timely diagnosis of pathogenic bacteria is a determining factor in minimizing the spreading of infectious diseases and enhancing the survival rates (Lazcka et al. 2007).

The most conventional pathogenic bacteria detection methods include: (i) Culture-based methods, (ii) immunological tests such as enzyme-linked immunosorbent assays (ELISAs) and (iii) molecular tests such as polymerase chain reaction (PCR) (Ahmed et al. 2014; Nurliyana et al. 2018; Rajapaksha et al. 2019). However, these conventional methods have several disadvantages in terms of complexity, lack of sensitivity, cost, long procedural times and requisite for specialized facilities and well trained users (Rajapaksha et al. 2019). Therefore, there is a crucial need for rapid, sensitive and low-cost bacterial detection methods with minimal sample preparation. This has led to extensive research in developing biosensors for bacterial detection in recent years (Lazcka et al. 2007) which can be divided into two main categories: (i) those detecting target bacterial components, such as DNA, RNA or enzyme (Anderson et al. 2013; Foudeh et al. 2014; Miranda et al. 2011; Paniel and Baudart 2013) which require sample processing to release the desired components and (ii) those detecting whole bacteria. The additional time and cost required for sample processing in the first category is a major drawback which makes whole bacteria detection a more desirable technique.

Optical biosensors are the most promising technologies for whole bacteria detection. The most commonly studied optical biosensors for whole bacteria detection include: (i) Fluorescence-based sensors which rely on secondary fluorescently labeled reagents binding to captured bacteria on the surface, (ii) Label-free sensors such as Surface Plasmon Resonance (SPR) which detect the refractive index change of the sensor surface due to the bacteria accumulation. However, the additional time and cost required for the fluorescent labeling step in fluorescence-based methods and the low sensitivity and considerable cost consumables for SPR techniques are major disadvantages of these biosensors which limit their applicability for rapid, cost-effective and sensitive detection of bacteria (Ahmed et al. 2014).

In this study, we present a sensitive, label-free and rapid whole *E. coli* bacteria detection and imaging technique based on our Interferometric Reflectance Imaging Sensor (IRIS). Earlier applications of IRIS included ensemble measurements of biolayer height increase due to the accumulated biomass on the functionalized sensor surface (Daaboul et al. 2011; Lopez et al. 2011). The second modality of this biosensor called Single Particle IRIS (SP-IRIS) is capable of detecting single biological particles such as whole virus (Daaboul et al. 2014; Scherr et al. 2016; Sevenler et al. 2018), exosomes (Daaboul et al. 2016) and proteins (Monroe et al. 2013). In this work, for the first time, we utilize SP-IRIS for label free detection of whole *E. coli* bacteria in buffer solution with a limit of detection of 2.2 CFU/ml with no sample preparation or modification. We also show the applicability of this biosensor for bacteria detection in real matrices such as tap water. Our simple, robust and rapid detection technique addresses the aforementioned limitations of the traditional bacteria detection techniques. The simple design of our optical setup with off the shelf optical components ensures a cost-effective instrument for bacteria detection without requiring specialized facilities or highly trained users. Moreover, the dual imaging modality of our biosensor offers both: (i) Rapid and high throughput bacteria detection enabled by the large field of view in low-magnification imaging modality and therefore rapid scan of the entire sensor area. (ii) Bacteria morphological characterization enabled by the high-magnification imaging modality which resolves down to single bacteria captured on the sensor surface. To the best of our knowledge, this is the first time the whole bacteria detection in a rapid and label-free manner have been reported with such low limit of detection of 2.2 CFU/ml.

## 2. Materials and Methods

**2.1. Preparation and Quantification of Bacterial Samples:** Genetically unmodified K-12 MG1655 (ATCC 47076) strain of *Escherichia coli* was purchased from ATCC. *E. coli* colonies were scraped from frozen stock and incubated overnight in Luria-Bertani (LB) broth at 37 ºC. After 14 hours of incubation, *E. coli* culture was spun down at 13,000 rpm and 4 ºC for 10 minutes and then reconstituted in sterile Phosphate Buffered Saline (PBS). The reconstituted culture solution was serially diluted in a 96-well plate and absorbance at 600 nm was measured using a spectrophotometer to determine the starting concentration of bacterial culture, as shown schematically in Figure 1b. The number of cells is directly proportional to the OD600 measurements (1 OD600 = $10^8$ CFU/mL). Tenfold serial dilution of the culture solution was performed to obtain sample concentrations ranging from $10^7$ cells/ml to 10 cells/ml, as described in Figure 1a. For the lower concentrations ($10^3$ cells/ml, $10^2$ cells/ml), concentration of bacteria sample was also quantified by plating 100 µl of culture solution into separate LB-agar plates, followed by incubation at 37 ºC for 12 hours and subsequently counting individual colonies on each plate. The images of these culture plates are shown in Figure 1c.

**2.2. Sensor Surface Functionalization:** The sensor chip consists of a layered Si/$SiO_2$ substrate which is modified with MCP-4 (copoly DMA-NAS-MAPS) (Cretich et al. 2004), a polymer typically used to coat glass, silicon, or other hydroxylated surfaces for microarray applications. The surface chemistry used to modify the substrate surface should meet the following requirements: 1- availability of functional groups for probe attachment, 2- preventing non-specific binding to the surface, 3- stability to environmental changes, 4- being ideally low cost, robust and easily prepared (Cretich et al. 2006). Therefore, MCP-4 copolymer from Lucidant Polymers (Sunnyvale, CA, USA) is used to immobilize antibodies for *E. coli* detection in this study which is recommended for DNA and peptide arrays. Each monomer forming the MCP-4 (copoly DMA-NAS-MAPS) copolymer (Cretich et al. 2004) has a different function: dimethylacrylamide (DMA) enables self-adsorption to the $SiO_2$ substrate, 3-(trimethoxysilyl)propyl methacrylate (MAPS) provides covalent binding to the substrate through silane functional groups, and acryloyloxysuccinimide (NAS) presents NHS ester groups which are utilized in biomolecule probe covalent immobilization. Therefore, MCP-4 is an ideal polymer for antibody immobilization on the substrate used in this study. Polymer coating was achieved through first, oxygen plasma treating the chips which forms hydroxylated surface to adhere the polymer to the substrate; and second, incubation of the chips in polymer solution for 30 minutes. After the incubation, the chips are rinsed with deionized water and kept under vacuum until used. The polymer coated chips can be stored under vacuum for up to 6 months. The polymer coating exhibits functional groups on the sensor surface which allows immobilization of antibodies of interest. The binding between the biomolecules and the solid surface during immobilization is strong enough to retain the molecules on the surface during the entire biosensing experiment. Moreover, the local chemical environment allows the immobilized molecules to retain a native conformation and functionality after immobilization.

**2.3. Sensor Preparation:** The antibody (Anti- *E. coli O* + *E. coli K* antibody), purchased from Abcam (Cambridge, MA, USA), was used to capture *E. coli* on the sensor at a concentration of 3.0 mg/ml. A Scienion sciFLEXARRAYER S3 (Berlin, Germany) spotter instrument was used to create an array of antibodies. This instrument provides a high precision and accuracy in spotting thousands of spots on the chip surface with a user defined pitch (300 µm in this study) between the spots and a droplet volume of ~200 pL. The periodicity of the spots and the humidity in the spotting chamber (set to 57%) were optimized in order to prevent merging of neighboring spots. The chips were left in the spotter chamber for 24 hours to allow time for the antibodies to immobilize on the surface. In addition to the antibody, bovine serum albumin (BSA at 1.0 mg/ml) was spotted on the same chips as a negative control in the experiments (Figure S1 in supporting information shows the image of a section of a spotted IRIS chip surface, used in this study). The chips were then blocked for 1 hour in a buffer of 0.1 M Tris and 50 mM Ethanolamine (pH of 9) and then washed with deionized water.

**2.4. Experimental Protocol:** The anti-*E. coli* antibody spotted IRIS chips are incubated in 1ml of *E. coli* dilutions in Phosphate-buffered saline (PBS) ranging from 10 CFU/ml to $10^6$ CFU/ml. As shown in Figure 1d the incubation is performed in a 24-wellplate on low speed shaker for 2 hours while the IRIS chip is resting at the bottom of each well at room temperature. In order to prevent surface-attached microbial agglomerations, we added Tween-20, a non-ionic surfactant, to each incubation well to a final concentration of 0.1%. Previous research has shown the role of surfactants in inhibiting biofilm formation in gram-negative bacteria; Wu et al. specifically demonstrated disruption of biofilm formed by *E. coli* at solid-liquid interface by Tween-20 (Mireles et al. 2001; Toutain-Kidd et al. 2009; Wu et al. 2013). During the incubation period, whole *E. coli* cells are captured on the IRIS chip surface by binding to corresponding surface-immobilized antibodies. The chips are then taken out of the incubation well and washed three times in 1x PBS solution for 1 minute on medium speed shaker before being air dried. The insets of Figure 1d show the 10mm x 10mm IRIS chips, and a zoomed-in image of the chip surface before incubation, which has been spotted with antibody for *E. coli* capture and BSA as a negative control. In addition, we performed control experiments to ensure that *E. coli* particles do not bind to the interior surface of the well plate during the 2-hour incubation time (described in detail in the Supporting information and Figure S2). In order to show the applicability of the proposed bacteria detection method in non-sterile, real matrices, we also performed the same experiment with tap water. To keep the bacteria intact and the antibody spots on the IRIS chips activated, the pH of the dilution sample is kept to 7.4 by preparing a base solution containing 9 parts tap water and 1 part 1x PBS. Next, the diluted tap water is spiked with *E. coli* samples with concentrations ranging from 10 to $10^7$ CFU/ml. The same incubation procedure is performed by adding 1ml of each dilution sample and 0.1% Tween-20 to each incubation well, placed at room temperature on medium speed shaker.

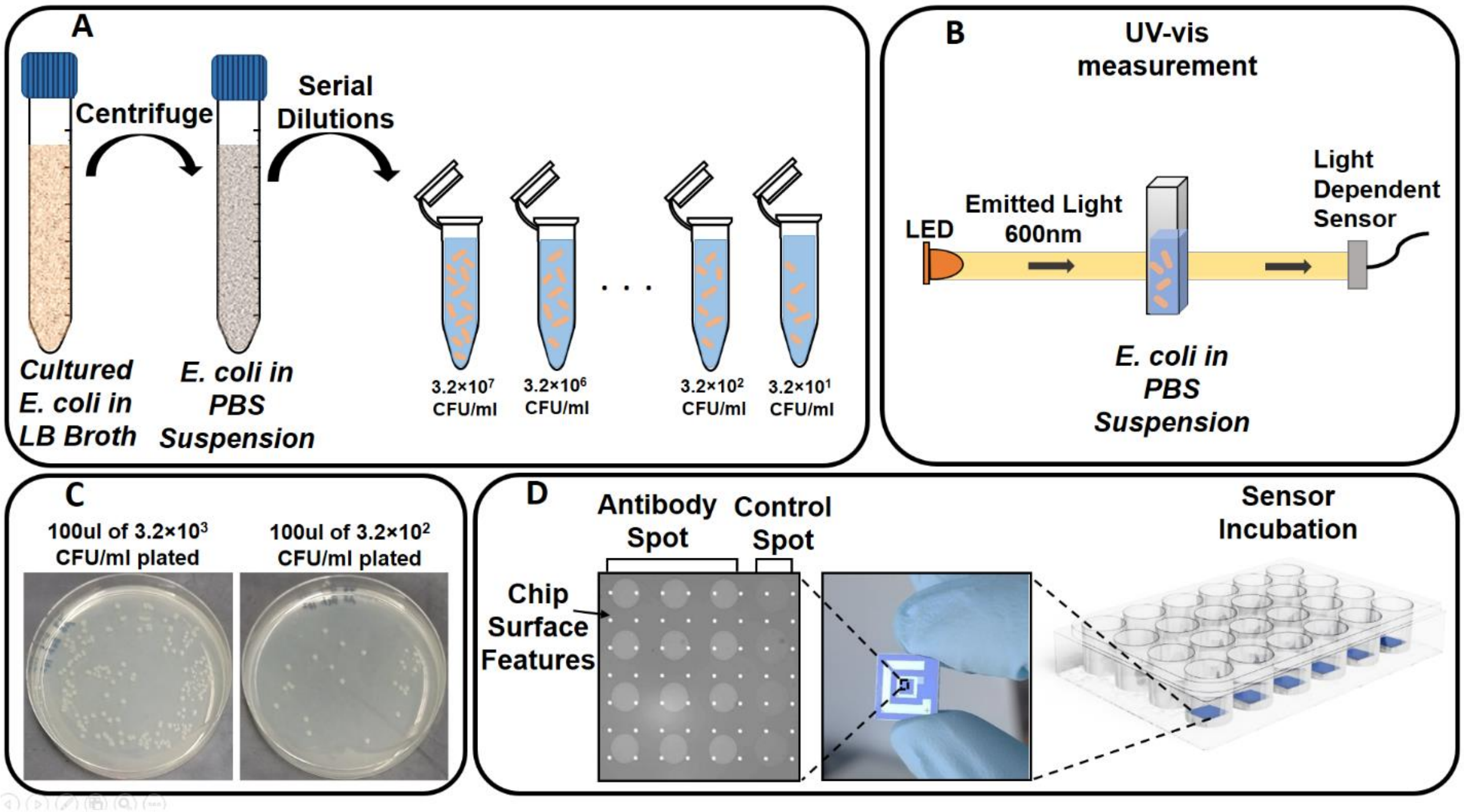

**Fig. 1. a.** Schematic of the bacterial sample preparation steps: centrifuging the *E. coli* in LB broth, reconstituting it in PBS and preparing the serial dilutions. **b.** Determining the original culture concentration by OD600 measurements using a spectrophotometer. **c.** Image of the culture plates from plating 100 µl of the two lowest concentrations of *E. coli* in PBS suspension. **d.** Experimental protocol steps: incubating the spotted IRIS chips in 1ml of *E. coli* dilutions in a well plate. The insets show the 10mm × 10mm IRIS chips, which has been spotted with *E. coli* antibody (3.0 mg/ml) and BSA (1.0 mg/ml) as negative control.

**2.5. The Optical Setup:** As shown schematically in Figure 2, the IRIS chip surface is illuminated through a Köhler illumination setup from an LED source at 520 nm wavelength. The integrating sphere connected to the LED source provides a uniform illumination on the sample. The illumination light is propagated by a 4f system to the beam splitter and the back focal plane of the objective. The light reflected from the sample is then imaged on to a CMOS camera through the tube lens of 200 mm focal length. A 5X magnification and 0.15 numerical aperture (NA) objective is used for imaging the chip surface. The CMOS camera has a 3.45 µm pixel size chosen to ensure a diffraction-limited system as described in equation (1) relating the minimum required pixel size (p) to the illumination wavelength ($\lambda$), magnification (M) and numerical aperture (NA) of the objective.

$$p \leq \frac{1}{2} \times M \times \frac{\lambda}{2 \times NA} \qquad (1)$$

We use a common path interferometric enhancement technique through a layered substrate, consisting of a thermally grown silicon dioxide layer on top of a silicon substrate, to enhance the visibility of the particles captured on the sensor surface, as shown in the zoomed-in view of the IRIS substrate (Avci et al. 2016). As described in previous publications (Avci et al. 2015), the constructive interference between the scattered light from the captured particles and the reflected light from the $Si/SiO_2$ interface, enhances the intensity of the light captured by the CMOS camera.

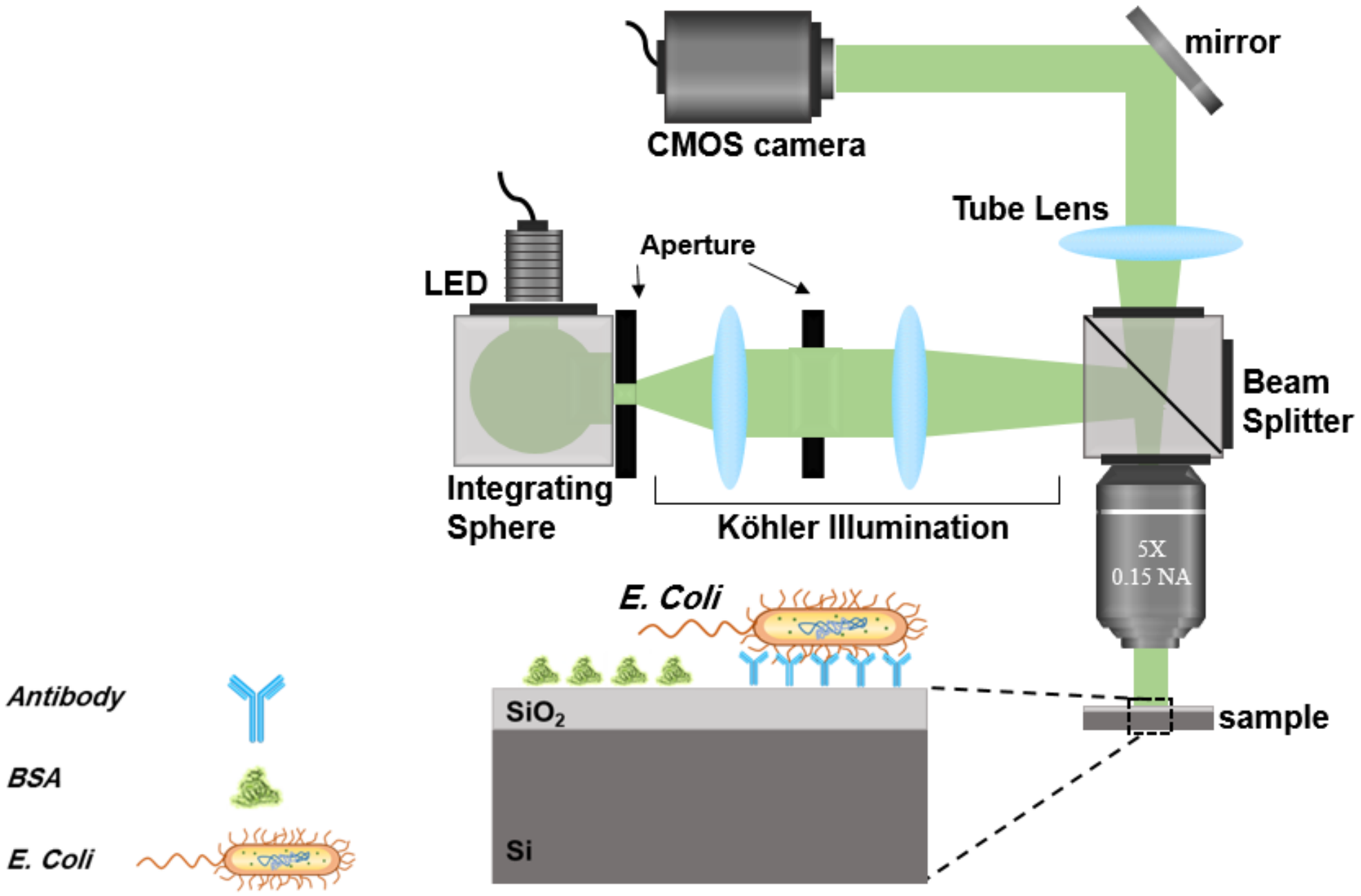

**Fig. 2.** Schematic of the optical setup used for *E. coli* detection. The 4f system propagates the LED light (at 525nm) to the beam splitter and the objective. The scattered light from the substrate is focused through the tube lens onto the CMOS camera. The zoomed-in inset shows the schematic of an IRIS chip consisting of layered $Si/SiO_2$ substrate with an *E. coli* bacterium captured on the antibody spotted region.

**2.6. Sample Imaging and Analysis Software:** The low-magnification modality of our biosensor provides imaging of the IRIS chips by a 5X magnification and 0.15 numerical aperture (NA) objective through the optical setup shown in Figure2a, at the visible wavelength of 520 nm. One of the most important advantages of this detection technique is the large field of view provided by the 5X objective, around 5.85 mm$^2$ (2.83 µm x 2.1 µm), which ensures a rapid scan of the entire IRIS chip for detection of the captured bacteria. We next perform a rapid digital detection and counting of the captured particles on the chip surface, using our custom-developed particle detection MATLAB algorithm, as shown schematically in Figure 3. The image of the sensor surface taken with the low-magnification modality is imported as the input of the algorithm, as shown in part (i). At the first step, the user enters the pixel values corresponding to: 1-The pixel coordinates of the left edge of first antibody spot on the top-left part of chip image and, 2- The pixel pitch between two adjacent spots. Using the entered parameters, the algorithm automatically registers and scans, one by one, the antibody and BSA spots on the chip, as shown by step 1 in Figure 3. Next, for each antibody and BSA spot, the algorithm should detect and count the captured bacteria particles which appear as a diffraction limited spot in the low-magnification images (Figure 3, part (ii)) and save the total counts in a final array. The basis of the particle detection algorithm is *Scale-Invariant Feature Transform (SIFT)* which requires optimizing the size and intensity parameters in order for the software to detect and count only the bacteria particles on the surface and prevent false positive counts. As shown in part (ii), in addition to bacteria particles there are salt residue from the dried PBS solution on the antibody spots which could cause a false positive in the final counts if the detection parameters are not well optimized.

In addition to the chemical specificity provided by the antibody spots to specifically capture the *E. coli* bacteria, shape and size characterization is also performed through the high-magnification imaging modality of our sensor, as shown in part (iii). We benchmarked the particle detection algorithm functionality against these high-magnification images where the rod-shaped bacteria are clearly resolvable, confirming that the parameters have been well optimized to count only the *E. coli* particles. Therefore, the parameter optimization is achieved through a closed loop feedback, steps 2 and 3 of Figure 3, which compares the detected particles in the low-magnification images to their corresponding high-magnification ones. The detection parameters' optimization is achieved by repeating the feedback loop only for a few antibody spots on each IRIS chip for different bacteria sample concentrations, therefore offering a rapid digital detection of the captured particles on our sensor surface. Refer to supporting information, Figure S3, for a more detailed explanation of the particle detection algorithm functionality.

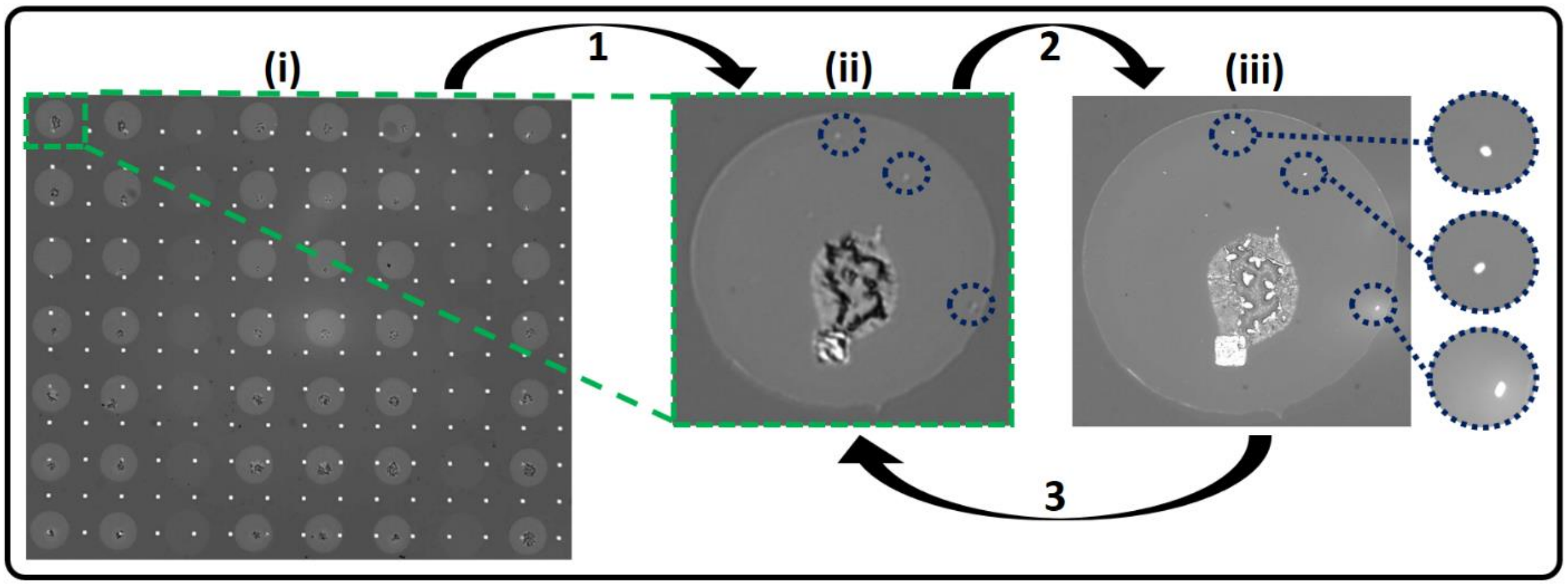

**Fig. 3.** Particle detection MATLAB algorithm step by step functionality process. (i). The IRIS chip surface image acquired by the low-magnification imaging modality, entered as the input of the algorithm. (ii). Zoomed-in view of one antibody spot in part (i), with the *E. coli* particles appearing as diffraction-limited spots shown in blue dashed circles. (iii). The high-magnification images of the same antibody spot shown in (ii), where the *E. coli* particles (shown in the blue dashed circles) appear as their original rod-shaped particles. The arrows labeled 1,2 and 3 correspond to steps: 1. Registering and scanning the antibody and BSA spots in each IRIS chip surface image. 2. Detection of the captured particles on the sensor surface in low-magnification images, through the initial size and intensity parameters entered by the user; and comparing them to the corresponding high-magnification images of the same antibody spot. 3. Feedback loop to step 2 for optimizing the detection parameters to count only the *E. coli* bacteria based on the size and shape of the captured particles in chip's high-magnification images.

## 3. Results and Discussion

**3.1. Statistical Analysis of The Sensor Response:** As mentioned above, the dual imaging modality of our optical setup allows for rapid scan and analysis of the entire chip surface while enabling morphological characterization and identification of the captured particles on the sensor surface, if so desired. Figures 4a and b show one field of view of the low-magnification imaging modality, for two IRIS chips incubated in medium (3.2 x $10^4$ CFU/ml) and low (3.2 x 10 CFU/ml) bacteria sample concentrations. The blue and green squares show two antibody spots which we have selected to perform high magnification characterization. The right panels of Figures 4a and b show a blow up of low-magnification images (the bottom row) of these two antibody spots, along with corresponding images acquired in high-magnification modality (the top row). Unlike the high-magnification images where one can easily resolve the single rod-shaped captured *E. coli* particles, they appear as diffraction-limited spots in the low-magnification modality images. Each antibody spot on the chip (Figure 4a), has at least a few captured *E. coli* particles. In contrast, in case of the lowest concentration, most of the antibody spots are blank as a direct consequence of the number of antibody spots on the chip (here 220) exceeding the total number of *E. coli* particles in the incubation sample. Figure 4c histogram plot indicates the frequency of antibody spots in one field of view of each IRIS chip based on the number of captured bacteria particles on each antibody spot for varying concentrations of bacteria in the target solution. An excellent fit to a Poisson distribution at each concentration confirms that bacteria capture represents independent events thus our detection modality would follow Poisson statistics. Therefore, a sampling of representative spots for verification through high-magnification imaging can lead to an accurate estimation of total number of captured bacteria hence the concentration in the target solution. For example, at low concentration, only 6 spots out of 50 have any potential detection events. At a sampling rate of 10%, one can validate nearly all detection events in high magnification allowing for an accurate estimation even in case of low number of events with the potential of false positives. In case of high concentrations, validating detection events on even fewer spots selected to represent the distribution will provide sufficient sampling for accurate estimation. Therefore, the burden of high-magnification imaging can be kept at a minimum (only 10% or less of the spots are re-imaged) regardless of the sample concentration and will improve the limit of detection by reducing the false positives at low concentration.

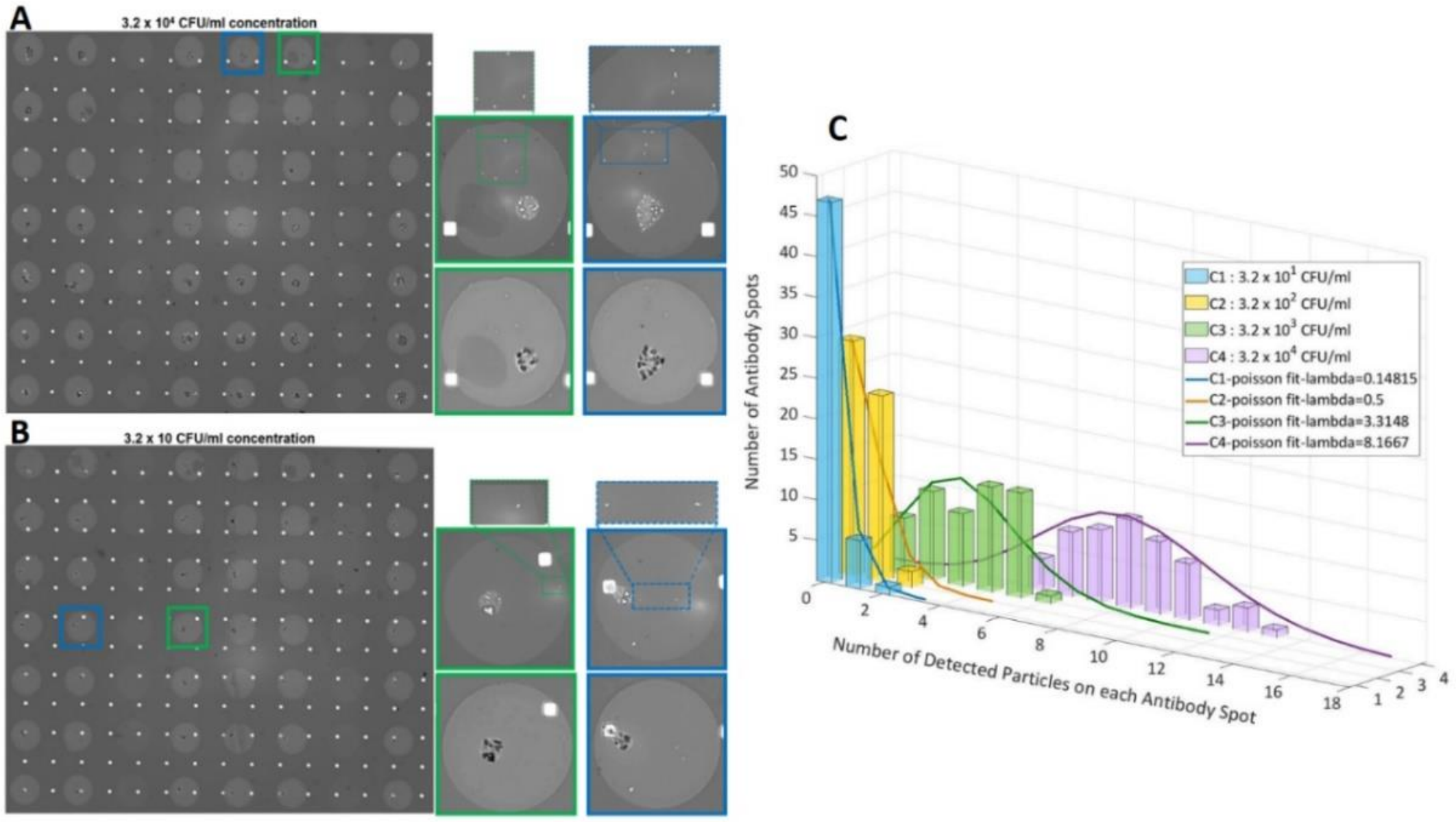

**Fig. 4. a-b.** Left panels show one field of view of the low-magnification imaging modality for two IRIS chips incubated in *E. coli* concentrations of 3.2 x $10^4$ CFU/ml and 3.2 x 10 CFU/ml, respectively. The blue and green squares show two sample antibody spots to perform shape and size characterization. The right panels show zoomed-in low-magnification images (the bottom row) of these two antibody spots and their corresponding high-magnification images (the top row) resolving rod-shaped *E. coli* bacterium. **c.** Histogram count of antibody spots on 4 different IRIS chips based on the number of captured bacteria particles on each antibody spot. The bar plots with colors blue, yellow, green and purple correspond respectively to *E. coli* sample concentrations of 3.2 x 10 CFU/ml to 3.2 x $10^4$ CFU/ml, shown with C1 to C4. The color-coded line plots show the Poisson distribution fits corresponding to each histogram count, confirming the independency of binding events from each other.

**3.2. Calibration Curve:** In order to obtain the calibration curve in our experiment for quantitative analysis of unknown sample concentrations, we incubated IRIS chips with *E. coli* in PBS dilutions for concentrations ranging from 3.2 x 10 to $10^6$ CFU/ml. Figure 5a shows a semi-log plot of the average particle count per $mm^2$ for both antibody and BSA spots as a function of *E. coli* concentration. An *E. coli* concentration of 0 in the plot corresponds to blank PBS sample incubation. The inset shows the zoomed-in plot for the first three concentrations for better comparison with the average counts of the negative control spot. Note that each antibody spot has a diameter of 160 µm, and therefore the area of 50 antibody spots corresponds to 1 $mm^2$ of a functionalized IRIS chip. To obtain the average particle count per $mm^2$, we first sum the particle counts on every 50 antibody spots and then obtain the average of the summations on each IRIS chip. The standard deviations of all the summation counts on each chip is shown as error bars for the corresponding sample concentration. As observed in this plot, there is a consistent increase in average particle counts for the immobilized antibody spots with an increase in *E. coli* concentration. However, the average count on the BSA control spots shows a very slight increase and stays relatively constant with an increase in *E. coli* sample concentration, confirming that we do not observe nonspecific binding of *E. coli* bacteria to the BSA control spots. As shown in Figure 5b, a desirable linear relationship was found in the log-log plot of the average particle count per $mm^2$ on antibody spots and the *E. coli* sample concentrations. The corresponding linear regression equations is $y = 0.46 x + 0.2521$ with a good correlation of $R^2 = 0.93$. Using the slope of the regression line and the sensor response to the blank sample, we calculate the limit of detection (LOD) per $mm^2$ (based on the formula of "Average count + 3*Standard Deviation of counts for Blank sample") to be 2.2 CFU/ml.

**3.3. Effect of Increasing the Sensor Area on the Assay Sensitivity:** In order to investigate the effect of sensor area on the sensitivity of our biosensor, we calculate the LOD per $mm^2$ for different sensor sizes, as shown in Figure 5c. Here, the sensor size is defined as the surface area of the immobilized antibody spots on the IRIS chip; therefore, each data point in this plot correspond to different sensor sizes, considered as the combination of 1, 4, 9, …, 50 antibody spots on the chip. Basically, the average and standard deviation of the particle count for combination of n spots is calculated for the chip incubated in the blank sample. Theoretically, from the LOD formula dependency on the standard Deviation, we expect a decrease in the LOD per $mm^2$ by increasing the sensor size, with a rate of $1/\sqrt{n}$ where n is the sensor area. As shown in Figure 5c, the experimental data points' decrease rate shows a close agreement to the expected theoretical value, confirmed by the great correlation of $R^2=0.99$ to the nonlinear fitting function of $y = 23.32 x^{-0.59} +1$.

**3.4. Sensor Specificity Test:** To evaluate the detection specificity of our biosensor, we performed the same experimental condition as *E. coli* for other non-target bacteria. The spotted IRIS chips with Anti-*E. coli* antibody were incubated with target bacteria *E. coli* sample and other non-target bacteria samples including *S. aureus, K. pneumonia* and *P. aeruginosa* at $10^4$ CFU/ml. After the 2-hour incubation, the chips are taken out, washed and air dried for imaging. Figure 5b shows the average particle count per antibody spot for different bacteria samples and the standard deviation of the particle counts is shown as the error bars in the plot. The results show that our biosensor has a much higher affinity to *E. coli* K12 and negligible cross-reactivity to *S. aureus*, *K. pneumonia* and *P. aeruginosa*. Therefore, our developed sensor platform provides a great specificity to the capture of target bacteria, confirming its application in multiplexed samples.

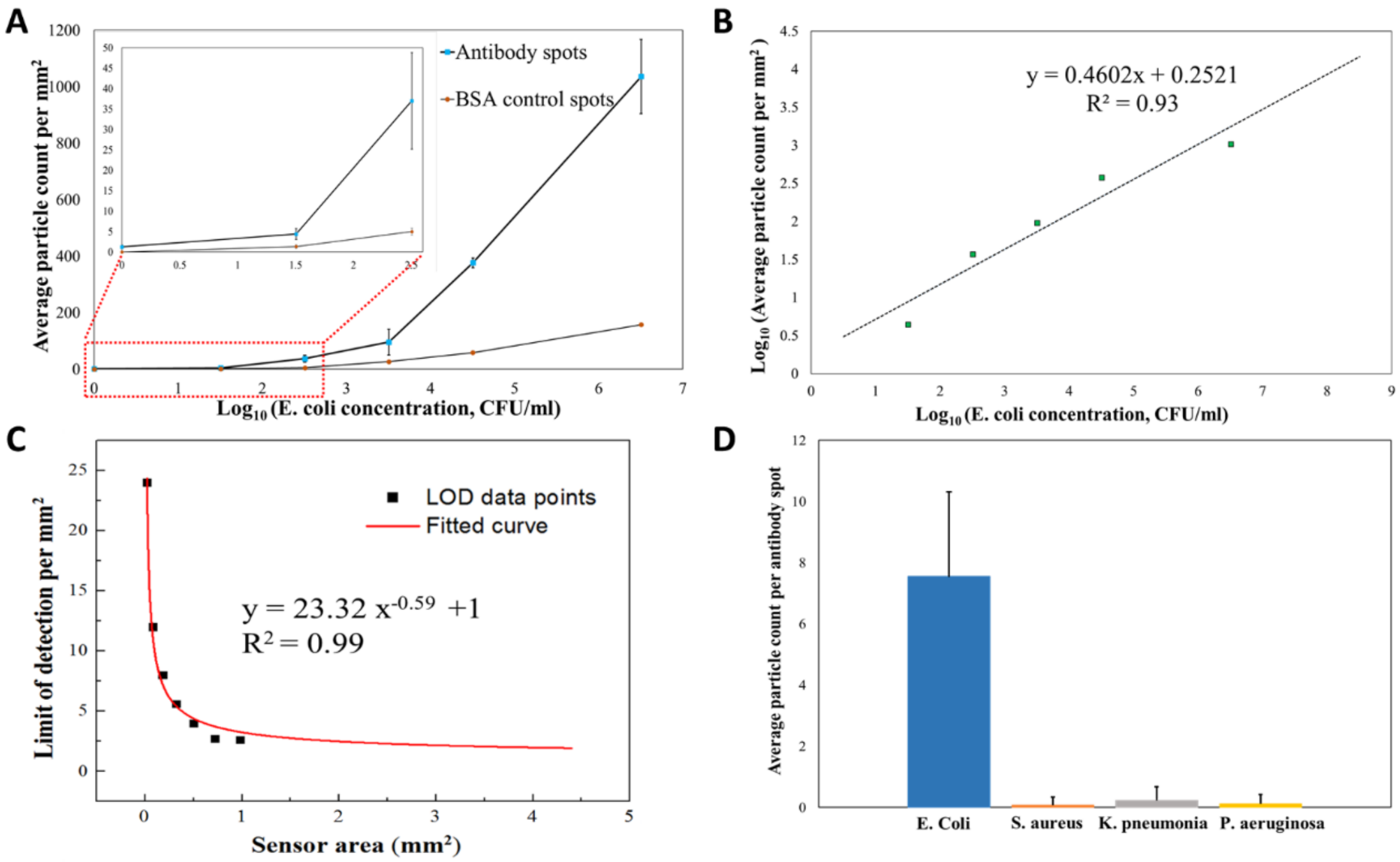

**Fig. 5. a.** Semi-log plot of the average particle counts per $mm^2$ for both antibody (blue squares) and BSA spots (orange circles) on the IRIS chips based on the *E. coli* solution concentration. The standard deviation of the counts is shown as an error bar for each concentration. The inset shows the zoomed-in section of the same plot for the blank sample and 3.2 ×10 and $10^2$ *E. coli* concentrations. **b.** log-log plot of the average particle counts per $mm^2$ based on the *E. coli* solution concentration for antibody spots (green squares). The dashed black line shows the linear fit to the data points, with a good correlation of $R^2 = 0.93$ to linear regression equation of y = 0.46 x + 0.2521. **c.** Plot of the limit of detection (LOD) per $mm^2$ based on different sensor sizes, shown with black square data points. The red solid line shows the non-linear fitting to the data points with the equation $y = 23.32x^{-0.59} + 1$ and a great correlation of $R^2 = 0.99$. **d.** Bar plot of average particle counts per antibody spot for spotted IRIS chips with Anti-*E. coli* antibody which were incubated in target bacteria *E. coli* sample and non-target bacteria samples including *S. aureus, K. pneumonia and P. aeruginosa* at $10^4$ CFU/ml. The standard deviation of the particle counts is shown as the error bars for each bacteria type.

**3.5. Tap water calibration curve:** In order to obtain the calibration curve for unprocessed tap water sample for quantitative analysis of unknown sample concentrations, we analyzed the images taken from the IRIS chips incubated in tap water spiked with *E. coli* samples with concentrations ranging from 3.23 x 10 to $10^7$ CFU/ml. The average particle counts per $mm^2$ for antibody and BSA spots as a function of *E. coli* concentration is shown by the semi-log plot in Figure 6a, where the 0 concentration corresponds to blank tap water incubation sample. A zoomed-in plot of the first three concentrations is also shown in the inset for a clearer comparison of the average counts. The antibody spots of the IRIS chips for tap water experiments have a diameter of 210 µm, and therefore the surface area of around 29 antibody spots corresponds to 1 $mm^2$ of a functionalized IRIS chip. The same analysis procedure of the PBS experiment is carried out here to obtain the tap water samples' calibration curve. As indicated in Figure 6a plot, unlike the BSA control spots, the average counts on the antibody spots show a clear increase by increasing the *E. coli* sample concentrations. Note that if we compare the calibration curves shown in Figures 5a and 6a, the average particle counts on the antibody spots for tap water is clearly lower than the PBS ones for the same *E. coli* concentrations. This lower count is due to the additional existing particles in tap water which are left on the chip surface, as shown in Figure S4 (will be added). Inevitably, some of these additional particles appear in the low-mag images with pixel size and intensity values which are close to the ones corresponding to smaller *E. coli* particles, making it challenging for the particle detection algorithm to differentiate them. So, we optimize the detection parameters such that it counts only the *E. coli* particles distinguishable from the existing particles in tap water to prevent false positive counts; therefore, excluding some *E. coli* particles from the final count. Figure 6b depicts the log-log plot of the antibody spots' average counts as a function of *E. coli* sample concentration which show a good correlation of $R^2 = 0.89$ to the linear regression line y = 0.34 x + 0.4136.

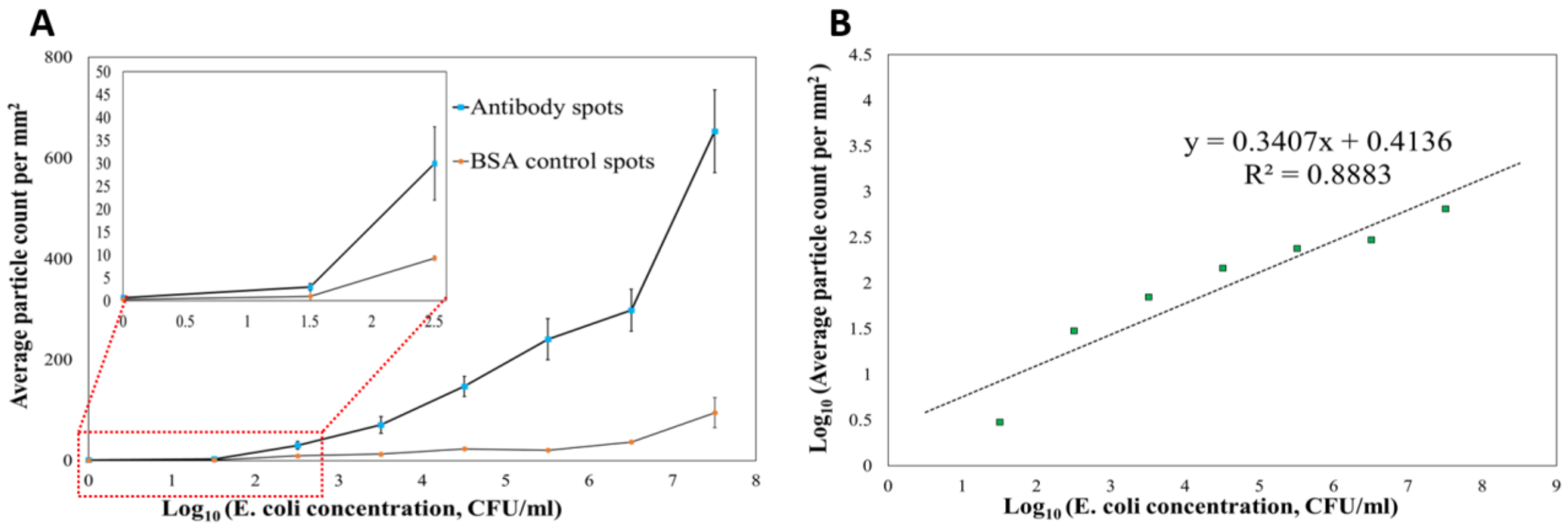

**Fig. 6. a.** Semi-log plot of the average particle counts per mm$^2$ for both antibody (blue squares) and BSA spots (orange circles) on the IRIS chips based on the *E. coli* spiked tap water concentration. The standard deviation of the counts is shown as an error bar for each concentration. The inset shows the zoomed-in section of the same plot for the blank sample and 3.23 ×10 and $10^2$ *E. coli* concentrations. **b.** log-log plot of the average particle counts per mm$^2$ based on the *E. coli* solution concentration for antibody spots (green squares). The dashed black line shows the linear fit to the data points, with a good correlation of $R^2 = 0.89$ to linear regression equation of y = 0.34 x + 0.4136.

Here, we successfully demonstrated the *E. coli* bacteria detection with an LOD as low as 2.2 CFU/ml, by digital detection and counting of the captured *E. coli* particle on the sensor surface through antibody probes. In addition to the conventional bacteria detection techniques, including culture and colony counting, ELISA, and PCR, several optical biosensing techniques have been developed in recent years to address the limitations associated with traditional techniques. Table 1 shows a comparison of our work with the most common optical biosensors in terms of their sensitivity, detection-time, bioreceptor and sensor complexity (Ahmed et al. 2014; Yoo and Lee 2016). Here, we focus on the biosensors which provide whole cell *E. coli* bacteria detection capability, therefore requiring no sample pre-processing which adds extra time and cost.

**Table 1. Comparison of current study with other biosensors for whole cell bacteria detection**

| Sensing Approach | Bioreceptor | Sensor assembly | LOD | Assay time | Ref |
|---|---|---|---|---|---|
| Fluorescence | Antibody | Antibody coated magnetic beads for capture, fluorescent polymeric micelles bioconjugated to antibody for response signal | 15 Cells/ml | ~ 1.5 hr | (Mouffouk et al. 2011) |
| Fluorescence | Probe-free | Fluorescently labeled bacteria detection by Microfluidic Device with a Hierarchical 3D Nanostructured detection platform | 50 CFU/ml | 1 hr | (Jalali et al. 2018) |
| Thin-film optical interference spectroscopy | Antibody | Direct cell capture on antibody-functionalized nanostructured oxidized porous silicon | $10^4$ Cells/ml | Several minutes | (Massad-Ivanir et al. 2011) |
| Long-range SPR | Antibody | Antibody coated magnetic nanoparticles for cell delivery, bacteria capture on antibody-functionalized SAM-gold surface | 50 CFU/ml | ~ 30 min | (Wang et al. 2012) |
| Long-period fiber gratings (LPFGs) | T4 Bacteriophage | Bacteriophage covalently immobilized on optical fiber surface | $10^3$ CFU/ml | ~ 20 min | (Tripathi et al. 2012) |
| SERS | Ag NP | Surface-enhanced Raman scattering (SERS) detection with Ag-NP coating the cell wall of bacteria | $2.5 \times 10^2$ CFU/ml | 10 min | (Zhou et al. 2014) |
| Colorimetry | Aptamer | Lateral flow (LF) test strips with DNA aptamers for capture, red-emitting quantum dots (Qdot 655) for response signal | 300–600 CFU/assay | ~ 20 min | (Bruno 2014) |
| Colorimetry | Aptazyme | Urease coupled to the DNAzyme on magnetic beads, and bacteria detection using litmus dye through pH increase of the reporting solution | $5\times10^2$ - $5\times10^3$ CFU/assay | 1-2 hr | (Tram et al. 2014) |
| Direct imaging through optical Interferometric enhancement | Antibody | Antibody-coated Si/$SiO_2$ substrate for direct capture and imaging of cells on substrate surface | 2.2 CFU/ml | 2 hr | Current study |

As indicated in Table 1, fluorescence-based biosensors could offer high detection sensitivity. However, the additional time and cost required for labeling the sample with fluorescent reagents, is a major drawback of this detection technique which hinders its applicability for a rapid and cost-effective biosensor (Ahmed et al. 2014). Recently, Müller et al. showed a smartphone-based fluorescence microscope which offers a cost-effectiveness and field-portable biosensor for bacteria detection in complex samples (Müller et al. 2018). SPR-based biosensors are one of the most common types of label-free optical biosensors. The early types of SPR-based biosensors suffered from low sensitivity, especially for detection of whole bacteria, due to similar refractive index of bacteria with the surrounding medium and also small penetration depth of the electromagnetic field into the bacteria. In recent years, various strategies have been introduced to improve the sensitivity level of SPR biosensors such as long-range SPR and SERS as shown in Table 1. However, SPR biosensors in general are complex, expensive and large equipment and therefore not suitable for a point of care diagnostic tool (Ahmed et al. 2014; Yoo and Lee 2016). To address these limitations, Tripathi et al. developed a more compact and cost-effective SPR biosensor based on long-period fiber gratings. Nonetheless, it suffers from low sensitivity as shown by its higher LOD compared to the other SPR biosensors. Finally, the colorimetric biosensors offer a portable, cost-effective detection technique which rely on color change of the reporting solution, detectable by naked eye due to the bacteria presence. However, these biosensors have several disadvantages including low sensitivity, limited multiplexing capability and limited quantification capability (Yoo and Lee 2016). Our biosensing platform offers an LOD of 2.2 CFU/ml with a 2-hour detection time for rapid and highly sensitive and specific bacteria detection through an interferometric substrate which can be functionalized with many capture probes for a high-throughput detection. The multiplexing capability of this sensor is also simply achieved by choosing different capture probes specific to different bacteria types. In addition, its simple, low-cost and label-free sensing modality addresses the limitations associated with the other optical biosensors. Note that the current LOD is achieved by particle counting only by rapid image acquisition at low optical magnification without a secondary verification at high magnification. Therefore, this LOD is partially constrained by the background counts that can be further reduced by the aforementioned verification step. By performing the verification step for every single bacteria capture event, the false positive counts can be completely eliminated leading to a zero background; thus, further improved LOD.

## 4. Conclusion

In summary, here we have presented a rapid, label-free and cost-effective optical biosensor for detection of *E. coli* with an LOD of 2.2 CFU/ml. Our approach relies on direct binding of whole bacteria to the sensor surface through antibody-antigen interaction, which is advantageous in terms of no sample pre-processing requirement, compared to the techniques detecting bacterial molecular components. Also, our detection modality follows the Poisson statistics due to independency of bacteria binding events, thus reducing the chances of false positive. This is achieved by validating the captured bacteria through high magnification imaging for the antibody spots which are best representative of the sample concentration. In addition, to the best of our knowledge, the detection sensitivity and rapidity offered by this biosensor exceeds the capability of other competing bacteria detection techniques. Although our endpoint measurement technique enables highly sensitive bacteria detection; a real-time detection technique would also provide valuable information on the bacteria binding kinetics such as association and dissociation rates. The future direction of this study aims at developing the real-time bacteria detection sensor with comparable sensitivity and rapidity to the current technique. The real-time detection modality will also decrease the detection time, since every bacteria binding event is visible by the diffraction limited spots in the large field of view; as opposed to waiting for the 2-hour incubation time before imaging and analyzing the data. In addition, we are investigating employing our proposed optical biosensor as a tool for monitoring the bacteria growth rate which is a critical tool in studying the bacteria resistivity to different types of antibiotics. This rapid technique could replace the time-consuming culture method which is the current clinical practice for bacteria resistivity studies. It has been also shown recently that using an interferometric imaging technique with appropriate image processing and analysis allows detecting and counting of colony forming units in a more rapid way compared to traditional colony counting techniques (Larimer et al. 2019). We believe the large field of view and real-time detection capability of our proposed biosensor, allows us to monitor the bacteria size growth continuously and recognize and count the colony forming units much sooner than the time it takes for the colony units to be visible to the naked eye in a traditional counting method.

## Credit authorship contribution statement

**Negin Zaraee:** Conceptualization, Investigation, Methodology, Data analysis, Visualization, Writing - original draft, Writing - review & editing. **Fulya Ekiz kanik:** Conceptualization, Investigation, Methodology, Writing - original draft, Writing - review & editing. **Abdul Bhuiya:** Investigation, Methodology, Writing - original draft, Writing - review & editing. **Emily Gong:** Investigation, Methodology. **M. T. Geib:** Methodology, Writing - review & editing. **Nese Lortlar Ünlü:** Supervision, Data curation, Writing - review & editing. **Ayca Yalcin Ozkumur:** Supervision, Validation, Data curation, Funding Acquisition, Writing - review & editing. **Julia Dupuis:** Supervision, Funding Acquisition, Project administration. **M. Selim Ünlü:** Supervision, Conceptualization, Funding Acquisition, Project administration, Validation, Data curation, Writing - review & editing.

## Declaration of competing interest

The authors declare that they have no known competing financial interests or personal relationships that could have appeared to influence the work reported in this paper.

## Acknowledgements

This work was financially supported by Army contract (grant number W56HZV-17-C-0182).